\documentclass[oupdraft]{sysbio}
\makeatletter
\def\ps@plain{\let\@oddhead\@empty\let\@evenhead\@empty
  \def\@oddfoot{\hfil\thepage\hfil}\let\@evenfoot\@oddfoot}
\let\ps@headings\ps@plain
\makeatother
\usepackage{url}

\usepackage{indentfirst}

\usepackage{booktabs}
\usepackage{siunitx}
\usepackage{multirow}
\usepackage{float}

\usepackage{wrapfig}
\usepackage{algorithm}
\usepackage[noend]{algpseudocode}
\usepackage{ragged2e}
\justifying
\usepackage{bm} 

\newcommand{\matQ}{\mathbf{Q}}
\newcommand{\matA}{\mathbf{A}}
\newcommand{\matB}{\mathbf{B}}
\newcommand{\matC}{\mathbf{C}}
\newcommand{\matD}{\mathbf{D}}
\newcommand{\matF}{\mathbf{F}}

\newcommand{\matP}{\mathbf{P}}
\newcommand{\matV}{\mathbf{V}}

\newcommand{\vecp}{\mathbf{p}}

\newcommand{\vecx}{\mathbf{x}}
\newcommand{\beast}{{\sc Beast X} }
\newcommand{\beagle}{{\sc Beagle} }
\newcommand{\numModules}{{M}}
\newcommand{\numStates}{{S}}
\newcommand{\numSites}{{L}}
\newcommand{\density}{{\delta}}
\newcommand{\numThreads}{{T}}
\newcommand{\numProbeCols}{{\ell_{\mathrm{est}}}}

\renewcommand{\theequation}{\arabic{equation}}
\renewcommand{\circ}{\odot}

\begin{document}
\nolinenumbers

\title{BEAGLE in Action: Phylogenetic Likelihoods for Very Large State Spaces}

\author{Benjamin D Redelings$^{1}$, April Wright$^2$, Karthik Gangavarapu$^3$, Yucai Shao$^4$, Filippo Monti$^4$, Wu-Min Deng$^5$, Shuo Su$^6$,  Andrew Rambaut$^7$, Guy Baele$^8$, Philippe Lemey$^8$, Marc A. Suchard$^9$ and Xiang Ji$^{10,\ast}$\\[4pt]
\textit{$^{1}$~Department of Mathematics, Tulane University, New Orleans, Louisiana, United States}\\
\textit{$^2$~Department of Biological Sciences, Southeastern Louisiana University, Hammond, LA}\\
\textit{$^{3}$~Department of Translational Medicine, The Scripps Research Institute, La Jolla, California, United States}\\
\textit{$^4$~Departments of Biostatistics, University of California, Los Angeles, California, United States}\\
\textit{$^{5}$~Department of Biochemistry and Molecular Biology, Tulane University, New Orleans, Louisiana, United States}\\
\textit{$^{6}$Shanghai Institute of Infectious Disease and Biosecurity, School of Public Health, Fudan University, Shanghai, China}\\
\textit{$^7$~Institute of Ecology and Evolution, University of Edinburgh, Edinburgh, UK}\\
\textit{$^8$~Department of Microbiology, Immunology and Transplantation, Rega Institute, KU Leuven, Leuven, Belgium}\\
\textit{$^9$~Departments of Biostatistics, Biomathematics and Human Genetics, University of California, Los Angeles, California, United States}\\
\textit{$^{10}$~Department of Statistics, Iowa State University, Ames, Iowa, United States}
\\[2pt]
\textit{*xiangji@iastate.edu}}

\markboth%
{Redelings et al.}
{Action}

\maketitle

\begin{abstract}
{Computing phylogenetic likelihoods becomes challenging when the continuous-time Markov chain has a large state space.
At each branch, pruning algorithms require terms of the form $e^{\matQ t}\cdot\matB$, where $\matQ$ is an $S \times S$ rate matrix (e.g.,~for $S$ states), $t$ is a branch length, and the columns of matrix $\matB$ (e.g.,~$\matB \in \mathbb{R}_{+}^{S \times L}$) are partial-likelihood vectors for $L$ site patterns.
Many implementations calculate the dense transition matrix $\matP(t)=e^{\matQ t}$ via eigendecomposition and multiply it by $\matB$.
This strategy is efficient for small state spaces, but it can be inefficient for large sparse rate matrices because forming $\matP(t)$ or its eigen vector matrices destroys sparsity and its cost may not be well amortized when $L$ is small.
We implement an alternative likelihood kernel based on the  \citet{al2011computing} algorithm for computing the action of the matrix exponential directly, avoiding explicit construction of $\matP(t)$ or its eigen vector matrices.
The resulting method takes full advantage of the sparsity of $\matQ$.
We implement this method as a CPU-based extension to the \beagle\ library and evaluate it on two large-state molecular models: a Markov-modulated codon model of heterotachy with $\numModules$ selective regimes and $61\cdot\numModules$ possible states, and a codon-pair model of interlocus gene conversion between two paralogs with $61^2=3\,721$ possible states.
Our results characterize when action-based likelihood computation improves performance over calculating $e^{\matQ t}$ directly.
We show that this method enables fast, scalable inference for complex models, removing a key computational barrier to the study of evolution with large state space models.}
{Phylogenetics, Likelihood, Matrix Exponential, BEAGLE, Large State Space.}
\end{abstract}
\newline

\section{Introduction}

Phylogenetic models often describe the evolution of discrete-state characters along an evolutionary tree.
The states commonly represent the 4 DNA nucleotides, the 20 amino acids, or the 61 non-stop codons.
However, many evolutionary questions require richer state spaces.
Hidden-state models augment the observed alphabet with unobserved rate or process categories \citep{tuffley1998modeling,huelsenbeck2002testing}.
Markov-modulated codon models allow the strength of selection, often measured by $dN/dS$, to change across branches or through time, yielding $61\cdot \numModules$ states for $\numModules$ selective regimes \citep{guindon2004modeling,baele2021markov,gangavarapu2024many}.
Polymorphism-aware models expand the nucleotide alphabet to represent population-level allele frequencies, allowing substitution, drift, and fixation to be modeled jointly \citep{de2015pomo,schrempf2016reversible,borges2022nucleotide}.
Models of coevolution between amino-acid sites can use state spaces of size $20^2=400$ \citep{yeang2007detecting,dib2014evolutionary}, and codon models with dependence between sites or loci can be much larger \citep{rodrigue2009computational}.
For example, interlocus gene conversion (IGC) between two paralogs can be modeled using codon-pair states, giving $\numStates=61^2=3\,721$ states \citep{ji2016phylogenetic, yang2023interlocus}.
Thus, large-state continuous-time Markov chains arise naturally in studies of molecular evolution and population genetics.

Large state spaces also arise in models of organism-level traits.
Unlike characters based on DNA or protein sequences, the number of observed characters may be small, sometimes only one or a few.
For example, models of geographic range evolution require up to $2^R$ states for $R$ geographic regions \citep{ree2008maximum}, but operate on just a single character.
Chromosome-number evolution requires a large count-valued state space to represent dysploidy and polyploidy events \citep{mayrose2010probabilistic,glick2015chromevol} and also operate on a single character.
Other examples include microsatellite repeat counts \citep{calabrese2005models,wu2011joint}, discretized continuous traits \citep{boucher2016inferring,boucher2018general}, and gene-family copy-number evolution \citep{deBie2006cafe}.
These models show that computational difficulty is not only a property of sequence alignments, and that large state spaces may occur precisely in settings where there are too few characters to amortize expensive transition-matrix calculations.

Computing likelihoods under these models is challenging because pruning algorithms repeatedly require computing the action of a branch transition matrix on a vector or matrix of partial likelihoods.
We write $\numStates$ for the number of continuous-time Markov chain (CTMC) states and $\numSites$ for the number of columns in the compressed alignment, equivalent to the number of unique site patterns, and the number of columns of $\matB$ where $\matB$ is a real matrix of size $S \times L$ (i.e.,~$\matB \in \mathbb{R}_{+}^{S \times L}$).
For a branch of length $t$ with rate matrix $\matQ \in \mathbb{R}^{S \times S}$, the transition probability matrix is
\begin{equation}
  \matP(t) = e^{\matQ t}.
\end{equation}
For eigendecomposition $(\matV,\matD)$ of $\matQ$ where $\matQ = \matV\cdot \matD\cdot \matV^{-1}$, the transition probability matrix becomes
\begin{equation}
	\matP(t) = \matV \cdot e^{\matD \cdot t} \cdot \matV^{-1},
\end{equation}
and may be kept as matrix products to reduce computations when multiplied by $\matB$.
If $\matA=\matQ t$, a pruning step requires an expression of the form
\begin{equation}
  e^{\matA}\cdot\matB.
\end{equation}
For sequence data, these columns usually correspond to sites or site patterns.
For organism-level traits, they may correspond to one or a few discrete characters.

Many phylogenetic likelihood implementations first compute the dense transition matrix $\matP(t)$ (with or without explicitly forming the matrix itself) and then multiply it by $\matB$.
This strategy is effective for small state spaces such as DNA and is most efficient when the cost of computing $\matP(t)$ can be amortized over many sites.
However, it can be extremely inefficient for large sparse rate matrices.
Even when $\matQ$ is sparse, the matrix exponential $e^{\matQ t}$ is generally dense, so explicitly forming $\matP(t)$ or its dense eigen vector matrices loses the sparsity of the underlying process.
The cost of this approach is roughly the cost of forming these dense matrices, often cubic in $\numStates$, plus the cost of multiplying the resulting dense matrix by $\matB$.
This leads to a cost of $O(\numStates^3)$ for eigendecomposition and $O(\numStates^2\numSites)$ for matrix-matrix multiplication.
For a data set with $N$ tips, the total cost amounts to $O(\numStates^3 + \numStates^2\numSites N)$ for homogeneous processes where one computes eigendecomposition just once per likelihood evaluation, and $O(\numStates^3 N + \numStates^2 \numSites N)$ for inhomogeneous processes where each branch requires one eigendecomposition calculation.
As $\numStates$ grows, this cost can become a substantial barrier to using biologically realistic models.

The relevant computational tradeoff depends not only on the number of states $\numStates$, but also on how many characters $\numSites$ we have.
When $\numSites \ll \numStates$, forming a full $\numStates \times \numStates$ transition matrix is especially wasteful because only its action on a few partial-likelihood vectors is needed.
This occurs in models of chromosome number, biogeographical range, etc., where $\numSites = 1$, and we only need the action on a single partial-likelihood vector.
At the other extreme, when $\numSites \gg \numStates$, the cost of forming $\matP(t)$ can be amortized over many site patterns, and optimized dense matrix multiplication may become more competitive.
This can occur in sequence data sets with long alignments, but it is important to keep in mind that the meaning of ``long'' depends on whether the data being analyzed is DNA ($\numStates=4$), amino acids ($\numStates=20$), codons ($\numStates=61$), or something with even larger $\numStates$.
In this regime, the benefits of computing the action instead of directly forming $\matP(t)$ depend strongly on the sparsity of $\matQ$ as well as on other factors such as the branch length and the relative efficiency of the dense versus sparse linear-algebra kernels.
We therefore seek to assess whether direct computation of $e^{\matA}\cdot\matB$ improves phylogenetic likelihood computation not only for one-character models, but also for sequence-like models with many sites when the rate matrix is sufficiently sparse.

The problem of computing matrix exponentials, and especially the action of the matrix exponential on a vector or matrix, has a long history in numerical linear algebra.
Krylov-subspace methods, as implemented for example in EXPOKIT, were developed to compute $e^{\matA}\cdot\vecx$ for large sparse matrices without explicitly forming $e^{\matA}$ \citep{sidje1998expokit}.
More recently, \citet{al2011computing} introduced an algorithm for computing $e^{\matA}\cdot\matB$ using scaling and truncated Taylor series approximations.
This method estimates the amount of scaling and the Taylor degree needed to meet a specified tolerance, and its dominant operations are products of $\matA$ with matrices having the same number of columns as $\matB$.
When $\matA$ is sparse, these products can be much cheaper than forming and applying a dense transition matrix.

Related ideas have already entered phylogenetics and phylogenetic-adjacent computation.
Lagrange-NG uses Krylov-subspace approximations to compute matrix-exponential actions, $e^{\matQ t}\vecp$, accelerating DEC likelihood calculations for historical biogeography models with large sparse range-state spaces \citep{bettisworth2023lagrange}.
The Al-Mohy--Higham action algorithm has also been used in phylodynamic inference, where \citet{leventhal2014using} used it to compute the likelihood of sampled transmission trees under a birth--death--sampling model while marginalizing over epidemic trajectories.
Uniformization provides another route around explicit matrix exponentiation by representing a CTMC as a discrete-time Markov chain observed at a Poisson number of potential jump times.
This idea has been used in phylogenetics for transition-probability calculations, data augmentation, and stochastic mapping \citep{mateiu2006inferring,deKoning2010rapid,irvahn2014phylogenetic}.
Other approaches reduce the effective state space itself, for example by aggregating states in codon- and amino-acid-pair models \citep{davydov2017state}.
These studies show that avoiding or modifying dense transition-matrix calculations can be valuable, especially for sparse CTMCs and models with few observed characters.
However, they do not provide a general likelihood kernel for large sparse molecular substitution models in which $\matB$ may contain anything from one site pattern to thousands of site patterns.

Here we implement direct computation of the matrix-exponential action as a CPU-based extension to the \beagle\ high-performance computing library for statistical phylogenetics \citep{ayres2012beagle, ayres2019beagle, gangavarapu2026beagle}.
Our implementation computes $e^{\matQ t}\cdot\matB$ without explicitly constructing $e^{\matQ t}$, using the Al-Mohy--Higham algorithm and sparse matrix--dense matrix products.
This preserves the sparsity of the rate matrix during the branch-propagation step and allows the same likelihood interface to be used for models with very large state spaces.
Because the advantage of the method depends on both the number of states and the number of site patterns, we compare the action-based kernel not only against the standard \beagle\ CPU implementation, but also against a branch of BEAGLE that optimizes dense matrix multiplication for large state spaces using the C++ library Eigen.
This comparison is important because inefficient dense multiplication can otherwise exaggerate the apparent benefit of the action method.

We evaluate the method on two large state space molecular models.
First, we examine Markov-modulated codon models in which the codon state is augmented by one of $\numModules$ selective regimes \citep{guindon2004modeling,baele2021markov,gangavarapu2024many}.
These models have $61\cdot \numModules$ states and provide a flexible framework for studying heterotachy in protein-coding evolution.
Second, we revisit the codon-pair model of IGC between two paralogs \citep{ji2016phylogenetic}.
In this model, the state at each site is a pair of codons $(x_i,y_i)$, giving $61^2=3\,721$ states.
Substitutions can occur in either paralog, and gene conversion events can replace one codon by the other, possibly changing multiple nucleotides simultaneously and tending to reduce the differences between the two paralogs.
Using these examples, we measure how runtime and accuracy depend on $\numStates$, $\numSites$, sparsity, and model structure.
Our results show that action-based likelihood computation can substantially expand the range of large sparse CTMC models that are practical for phylogenetic inference.

\section{New Approach}

To describe our new approach, we begin by introducing notation and describing the stochastic process.
We then describe how to compute the likelihood using the action of the matrix exponential.
With this framework in place, we describe the algorithm for computing the action when the Taylor series approximation order for $\matQ\cdot t$ is known, and how to compute the approximation order.
We propose an updated method for computing the necessary order for the approximation.
Finally, we describe our strategy for parallelization using multiple threads.
Readers primarily interested in the empirical performance of the method can focus on the notation, pruning recursion, and high-level action algorithm, and then proceed to the Results.
The subsections on choosing Taylor parameters, shifting, and parallelization provide additional mathematical and implementation detail for readers who want a deeper understanding of the algorithm.

\subsection{Notation and Stochastic Process}
We denote the rooted bifurcating phylogeny by $\mathcal{F}$. 
The tree has $N$ tips.
Tip nodes are indexed by $1,\ldots,N$, and internal nodes are indexed by $N+1,\ldots,2N-1$, with the root at node $2N-1$.
Each non-root node $i=1,\ldots,2N-2$ has a branch length $b_i$ that is associated with the rootward branch of node $i$.

Each compressed alignment column $c=1,\ldots,\numSites$ is assigned an independent and identically distributed (i.i.d.) rate class $R_c \in \{1,\ldots,\mathcal{R}\}$. 
The rate class is drawn from a mixture distribution with weights $\Pr(R_c=r)$, where each class $r$ corresponds to a scaling factor $\gamma_r$.

Character states are denoted $\mathbf{Y}=\{Y^c_i\}$, where $c=1,\ldots,\numSites$ indexes the compressed alignment columns and $i=1,\ldots,2N-1$ corresponds to the node of the tree.
If a site pattern represents multiple raw alignment columns, its multiplicity is applied when accumulating the log likelihood; this multiplicity does not affect branch propagation and is suppressed below.
States take values in $\{0,\ldots,\numStates-1\}$. 
The stochastic model treats the underlying alignment columns as i.i.d., and within each column the root state $\mathbf{Y}^{c}_{2N - 1}$ is drawn from the root state distribution $\pi$, often chosen as the stationary distribution, so that $\Pr(\mathbf{Y}^{c}_{2N - 1}=s)=\pi_s$.
Given $R_c=r$, characters in column $c$ evolve along each branch $i$ under the scaled rate matrix
\begin{equation}
   \matQ^{(r)}_i = \matQ_i \cdot \gamma_{r}.
\end{equation}
Evolution on different branches is conditionally independent given the state at the rootward endpoint of each branch. 

For each branch $i$ and rate class $r$, the transition probability matrix is
\begin{equation}
   \matP^{(r)}_i = e^{\matQ^{(r)}_i b_i}.
\end{equation}
Finally, we use the shorthand $\Xi(\matA,\matB)$ to denote the action of a matrix exponential on a matrix or vector, i.e.,
\begin{equation}
   \Xi(\matA,\matB) = e^{\matA}\cdot\matB.
\end{equation}

\subsection{Phylogenetic Action - Modifying the Pruning Algorithm}

We now modify the pruning algorithm to propagate conditional likelihoods along each branch without computing $e^{\matQ_i\cdot b_i}$ directly.
Suppressing site-pattern multiplicities in the notation, we have
\begin{equation}
\begin{aligned}
  \Pr(\mathbf{Y}) &= \prod_{c=1}^{\numSites} \Pr(\mathbf{Y}^c) \\
  \Pr(\mathbf{Y}) &= \prod_{c=1}^{\numSites} \left( \sum_r  \Pr(\mathbf{Y}^c|R_c=r)\cdot\Pr(R_c=r)\right ).
\end{aligned}
\end{equation}
The pruning algorithm \citep{felsenstein1981evolutionary} for computing this likelihood is a recursive algorithm that requires computing the partial likelihoods $p^{crs}_{k}= \Pr(\mathbf{Y}^c_{\lfloor k \rfloor}|Y^c_k=s,R_c=r) $ for each internal node $k$.
Here $\mathbf{Y}^c_{\lfloor k \rfloor}$ denotes the observed characters at the leaves of the subtree rooted at $k$.
If node $k$ has children $i$ and $j$, then we have:
\begin{equation}
\begin{aligned}
  \vecp^{cr}_k & = \matP^{(r)}_i \vecp^{cr}_i \circ \matP^{(r)}_j \vecp^{cr}_j \\
   & = (e^{\matQ^{(r)}_i\cdot b_i})\vecp^{cr}_i \circ (e^{\matQ^{(r)}_j \cdot b_j})\vecp^{cr}_j \\
   & = \Xi(\matQ_{i}\cdot \gamma_r b_{i}, \vecp^{cr}_i) \circ \Xi(\matQ_{j}\cdot \gamma_r b_{j}, \vecp^{cr}_j)
\end{aligned}
\end{equation}
where $\circ$ indicates the element-wise product of two vectors.
Thus, partial likelihoods are propagated along branches using the action of the matrix exponential instead of via multiplication by $\matP$.

\subsection{Action}
We now describe how to compute the action of $\matA$ on $\matB$ (i.e.,~$\Xi(\matA, \matB) = e^{\matA}\matB$).%
This description is intended to provide an understandable overview of how the algorithm works.
A reader who is intending to implement the algorithm will still need to read the original papers by \citet{al2011computing} and \citet{higham2000block} for the more subtle details.

The method of \citet{al2011computing} decomposes the action of $e^{\matA}$ into $s$ successive actions of $e^{\matA/s}$.
That is, after first applying $e^{\matA/s}$ to the input matrix $\matB$, the same operator is applied to the result, and so on for a total of $s$ applications:
\begin{equation}
\begin{aligned}
e^{\matA}\cdot \matB & =\underbrace{e^{\matA/s}\cdot e^{\matA/s}\cdot \ldots \cdot e^{\matA/s}}_{s\,\text{terms}}\cdot \matB\\
 & =\underbrace{e^{\matA/s}\cdot e^{\matA/s}\cdot \ldots}_{s-1\,\text{terms}}\cdot(e^{\matA/s}\cdot \matB)
\label{eq:action-scaling}
\end{aligned}
\end{equation}
The action of $e^{\matA/s}$ on $\matB$ is then approximated using a truncated Taylor series of order $m$:
\begin{equation}
\begin{aligned}
  e^{\matA/s}\cdot \matB & \approx\left(I+\frac{\left(\matA/s\right)}{1!}+\frac{\left(\matA/s\right)^{2}}{2!}+\frac{\left(\matA/s\right)^{3}}{3!}+\ldots+\frac{\left(\matA/s\right)^{m}}{m!}\right)\cdot \matB \\
   & \approx \matB+\frac{\left(\matA/s\right)}{1!}\matB+\frac{\left(\matA/s\right)^{2}}{2!}\matB+\frac{\left(\matA/s\right)^{3}}{3!}\matB+\ldots+\frac{\left(\matA/s\right)^{m}}{m!}\matB
\label{eq:taylor-action}
\end{aligned}
\end{equation}
The key breakthrough of \citet{al2011computing} is to efficiently determine values of $m$ and $s$ that achieve the desired accuracy with the least computing effort.

Each Taylor series term $\frac{(\matA/s)^j}{j!}\matB$ can be computed by multiplying the previous Taylor series term $\frac{(\matA/s)^{j-1}}{(j-1)!}\matB$ by $\frac{\matA}{s\cdot j}$ on the left.
Therefore each $m$th order expansion involves only $m$ matrix multiplications (Equation~\ref{eq:taylor-action}).
We must do this $s$ times, since the Taylor series computes one action of $\matA/s$ rather than the full action of $\matA$ (Equation~\ref{eq:action-scaling}).
This enables an efficient algorithm that requires $s\cdot m$ matrix multiplications.
Algorithm~\ref{alg:simple_action} below summarizes a slightly simplified version of Algorithm 3.2 in \citet{al2011computing}.
This skeleton omits several implementation details discussed by \citeauthor{al2011computing}, including the shifting and early-termination steps used in our implementation as well as the optional balancing step.

\begin{algorithm}
  \caption{- Abstract skeleton of Algorithm 3.2 in \citeauthor{al2011computing}}
  \label{alg:simple_action}
  \begin{algorithmic}
    \Function{ComputeAction}{$\matA, \matB_0, m, s$}
    \State $\matC \gets \matB_0$
    \For{$i \textbf{ in } 1:s$}
    \State $\matF \gets \matC$
    \State $\matB \gets \matC$
    \For{$j \textbf{ in } 1:m$}
    \State $\matB \gets \matA \matB / (sj)$
    \State $\matF \gets \matF + \matB$
    \EndFor
    \State $\matC \gets \matF$
    \EndFor
    \State{\Return $\matC \approx e^{\matA}\cdot\matB_0$}
    \EndFunction
  \end{algorithmic}
\end{algorithm}

Each multiplication by $\matA$ multiplies an $\numStates \times \numStates$ matrix by an $\numStates \times \numSites$ dense matrix, where the $\numSites$ columns contain the partial-likelihood vectors for the site patterns being propagated.
Algorithm \ref{alg:simple_action} therefore takes $O(s\cdot m\cdot \numStates^2 \cdot \numSites)$ computation time when $\matA$ is dense (i.e.,~every entry of $\matA$ is visited).
However, if $\matA$ is sparse, then the cost is proportional to the number of non-zero elements in $\matA$ (i.e.,~only non-zero entries in $\matA$ are visited), although the proportionality constant is higher.
Thus if a fraction $\density$ of the $\numStates^2$ elements of $\matA$ are non-zero, then the cost is $O(s\cdot m\cdot \density\cdot\numStates^2\cdot\numSites)$.
If $\density\propto 1/\numStates$, then the cost would be around $O(s\cdot m\cdot \numStates\cdot \numSites)$, which constitutes a substantial speedup.
However, the higher cost per non-zero entry can outweigh the savings from skipping zero entries.
In compressed sparse row (CSR) multiplication, the non-zero values and their column indices must be read, and the indices determine which rows of $\matB$ are accessed; the resulting access pattern can limit cache reuse \citep{bi2026memoryaware}.
In one CPU comparison, CSR multiplication became faster than dense multiplication at around 10\% matrix density \citep{heddes2025alwayssparse}.
It is important to note that the matrix powers $\matA^j$ are not directly computed.
Computing $\matA \cdot \matA^{j-1}$ via dense matrix multiplication takes $O(\numStates^3)$ computation time.
Additionally, $\matA^m$ is generally not sparse even when $\matA$ is sparse such that the above approach would still require close to $O(s\cdot m\cdot \numStates^3)$ computation time if we computed matrix powers directly.
Computation of $\matA^j$ is therefore not a viable option if we seek to avoid terms that are cubic in the number of states $\numStates$.
Algorithm \ref{alg:simple_action} therefore avoids directly forming the large, dense matrices $\matA^m$ in favor of repeated multiplication of a sparse matrix by a dense matrix.

\subsection{Determining $m$ and $s$}

The method of \citet{al2011computing} seeks to find the smallest value of $m\cdot s$ that yields sufficient accuracy.
The Taylor degree $m$ controls how many terms are used in each truncated Taylor expansion, while the scaling factor $s$ controls how many smaller exponential actions are composed to obtain the full action.
Larger values of either parameter increase the number of sparse matrix--dense matrix multiplications, so choosing them well is important for performance.
For a fixed matrix norm, increasing the Taylor degree $m$ increases the range of matrices that can be handled without scaling, so the parameter search tends to reduce $s$ by increasing $m$ first.
The upper bound on $m$ in the Al-Mohy--Higham algorithm prevents this from continuing indefinitely for numerical-stability reasons; consequently, $s$ tends to increase only after the required accuracy cannot be achieved within the allowed range of $m$.

The algorithm to compute $m$ and $s$ is specified in Code Fragment 3.1 of \citet{al2011computing}.
It evaluates bounds using quantities of the form $\alpha_p(\matA)=\max(d_p(\matA),d_{p+1}(\matA))$, where $d_p(\matA)=||\matA^p||^{1/p}$. %
With the typical choice $p_{\max}=8$, this can require estimates through $d_9$.
When $d_p(\matA) < ||\matA||$, the use of $d_p(\matA)$ rather than $||\matA||$ enables reaching the desired level of approximation error with smaller values of $m$ and $s$, and thus less computation.

However, as mentioned above, directly forming $\matA^p$ in order to compute $d_p(\matA)$ would be $O(\numStates^3)$, and thus would be too computationally expensive.
\citet{al2011computing} therefore approximate the norm $||\matA^p||$ using the method of \citet{higham2000block}.
This method approximates the matrix norm by computing $||\matA^p\cdot \mathbf{x}|| $ for specially chosen unit vectors $\mathbf{x}$.
$\matA^p\cdot\mathbf{x}$ can be computed via $O(p)$ multiplications of the sparse matrix $\matA$ times the dense vector $\mathbf{x}$, and thus has a cost that is $O(p\cdot \density\cdot\numStates^2)$.

\subsubsection{Deciding when to compute $d_p$}
In some cases the cost of computing the $d_p$ values for $p>1$ can exceed the benefit that they bring by decreasing $m$ and $s$.
\citet{al2011computing} therefore include an off-ramp to avoid computing the $d_p$ values in certain cases.
Their equation 3.13 sets out a condition under which the cost of computing the $d_p$ values would be more expensive than simply computing the action using the values of $m$ and $s$ obtained from $d_1$.
When this condition is true, their algorithm does not compute the $d_p$ values for $p>1$.
This condition is unlikely to be satisfied when the number of site patterns is large, but for small numbers of site patterns it can indeed be relevant.

However, we note that the $d_p$ have the largest effect for small values of $p$ (see Figure \ref{dvaluesIGC}).
These are also the cheapest $d_p$ values to compute.
We therefore modify the algorithm to compute $d_p$ values incrementally and on-demand during the computation of $m$ and $s$, saving the $d_p$ values for later use after they are computed.
We iterate over $p=2\ldots p_{\max}$, stopping when the amount of work done exceeds the cost of computing $e^{\matA}\matB$ under the $m$ and $s$ values based on all $p$ values that have been considered so far.
This approach has the benefit that when we determine whether computing an additional $d_p$ value is justified, we can use more realistic values of $m$ and $s$ that are based on higher $p$ values than $p=1$.
When the algorithm attempts to compute $m$ and $s$ for the same $\matQ$ later, it may then use the previously computed $d_p$ values for free, and compute $d_p$ values for higher values of $p$ if the amount of work is not too high.
We thus no longer need to depend on \citet{al2011computing} equation 3.13.
We additionally modify the cost estimates to use the number of sites per thread $\left\lceil \frac{\numSites}{\numThreads} \right\rceil$, where $\numThreads$ is the number of threads.
Let $\numProbeCols$ denote the number of probe columns used by the block 1-norm estimator.
This leads to the modified algorithm for choosing $m$ and $s$ given by Algorithm \ref{alg:choose_m_s}.
\begin{algorithm}
  \caption{- Selection of Taylor degree $m$ and scaling factor $s$}
  \label{alg:choose_m_s}
  \begin{algorithmic}
    \Function{ChooseMS}{$\matQ, t$}
    \State $(m^*,s^*) \gets (\infty,\infty)$

    \For{$m \textbf{ in } 1:m_{\max}$}%
      \State $s \gets \left\lceil ||\matQ \cdot t|| / \theta_m \right\rceil$%
      \State $(m^*, s^*) \gets (m,s)$ \textbf{if} $m \cdot s < m^* \cdot s^*$ 
    \EndFor

    \State $work \gets 0$

    \For{$p \textbf{ in } 2:p_{\max}$}
      \State $work \mathrel{+}= 4\numProbeCols\,p$ \textbf{if} \textbf{not} \Call{HasDValue}{$p$}
      \State $work \mathrel{+}= 4\numProbeCols\,(p+1)$ \textbf{if} \textbf{not} \Call{HasDValue}{$p+1$}

      \State \textbf{break} \textbf{if} $m^* \cdot s^* \cdot \left\lceil \numSites / \numThreads \right\rceil < work$

      \For{$m \textbf{ in } p(p-1)-1:m_{\max}$}
          \State $\alpha \gets \max(\Call{GetDValue}{p},\Call{GetDValue}{p+1}) \cdot t$
          \State $s \gets \left\lceil \alpha / \theta_m \right\rceil$

          \State $(m^*, s^*) \gets (m,s)$ \textbf{if} $m \cdot s < m^* \cdot s^*$ 
      \EndFor
    \EndFor

    \State{\Return $(m^*,s^*)$}
    \EndFunction
  \end{algorithmic}
\end{algorithm}
\subsection{Shifting}
\citeauthor{al2011computing} note that if we define the shifted matrix $\widetilde{\matA} = \matA - \mu\cdot \mathbf{I}$, we can then express the action of the matrix exponential in terms of $\widetilde{\matA}$:
\begin{align*}
  e^\frac{\matA}{s}\matB & = e^\frac{\widetilde{\matA}+\mu\mathbf{I}}{s}\matB \\
                        & = e^\frac{\mu\mathbf{I}}{s}e^\frac{\widetilde{\matA}}{s}\matB \\
                        & = e^\frac{\mu}{s}(e^\frac{\widetilde{\matA}}{s}\matB)
\end{align*}
We can thus compute the Taylor series using $\widetilde{\matA}/s$ and multiply by the constant factor $e^\frac{\mu}{s}$ afterwards.
If $d_p(\widetilde{\matA})$ is smaller than $d_p(\matA)$, this can result in smaller values of $m$ and $s$.
\citeauthor{al2011computing} recommend choosing $\mu$ so that the trace of $\widetilde{\matA}$ is zero, i.e.,~setting $\mu=\operatorname{trace}(\matA)/\numStates$.
They also describe optional balancing as a preprocessing step, but we do not use balancing in the benchmarks below.

\begin{figure}
  \begin{centering}
  \includegraphics[height=0.45\textwidth]{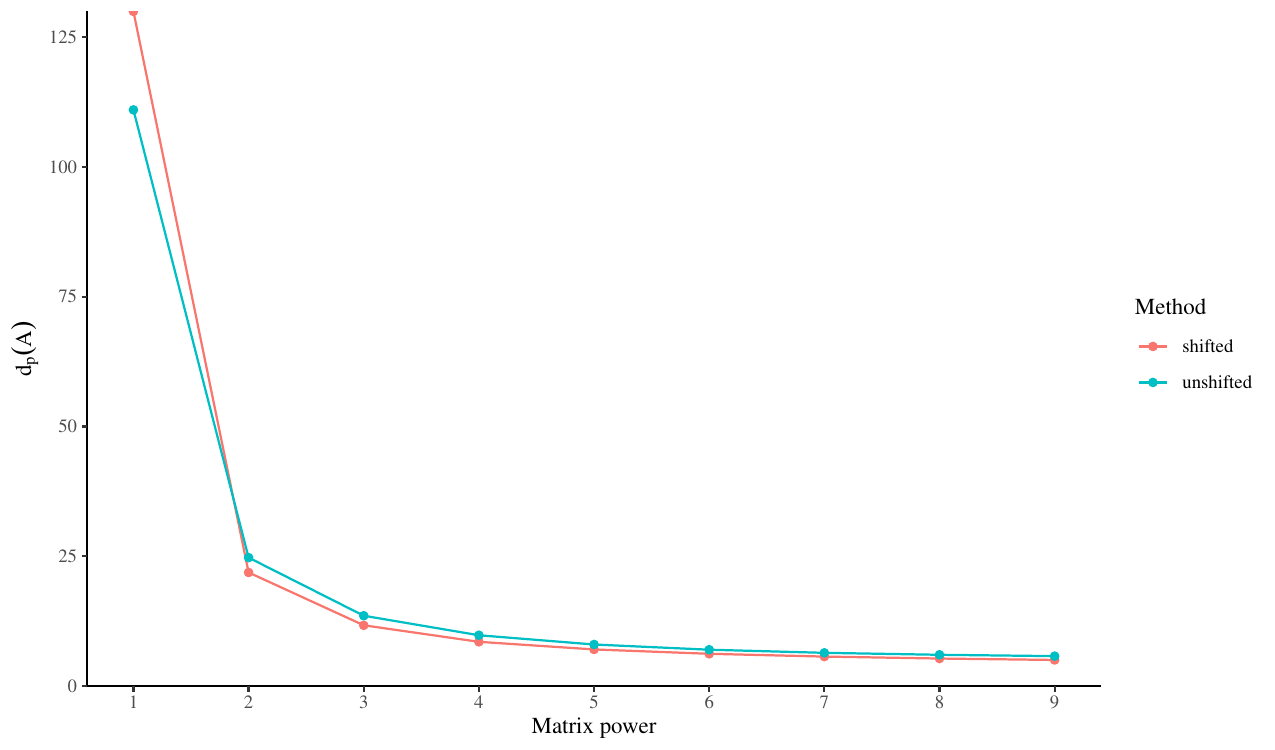}
  \par\end{centering}
\caption{\label{dvaluesIGC}Estimated $d_p(\matA)=||\matA^p||^{1/p}$ values for the IGC two-paralog model, with and without shifting.
Values are averaged over $\matA$ matrices encountered in a short MCMC analysis.
The rapid decrease from $p=1$ to larger $p$ explains why using higher powers can reduce the Taylor degree or scaling factor selected by the action algorithm.}
\end{figure}

We investigated how shifting $\matA$ in this way affected the $d_p$ values for the IGC and Markov-modulated examples analyzed in this study. %
We report $d_p$ values that are averaged over the course of a short Markov chain Monte Carlo (MCMC) analysis.
In all cases the computed $d_p$ values decrease monotonically with $p$, as expected.
For the IGC two-paralog model, the $d_p$ values decrease substantially as $p$ increases, from
$d_1=111.0$ to $d_9=5.74$.
Shifting did not substantially change these values, although it slightly increased $d_1$ and slightly decreased $d_p$ for $p>1$ (Figure~\ref{dvaluesIGC}).

\begin{figure}
  \begin{centering}
  \includegraphics[height=0.45\textwidth]{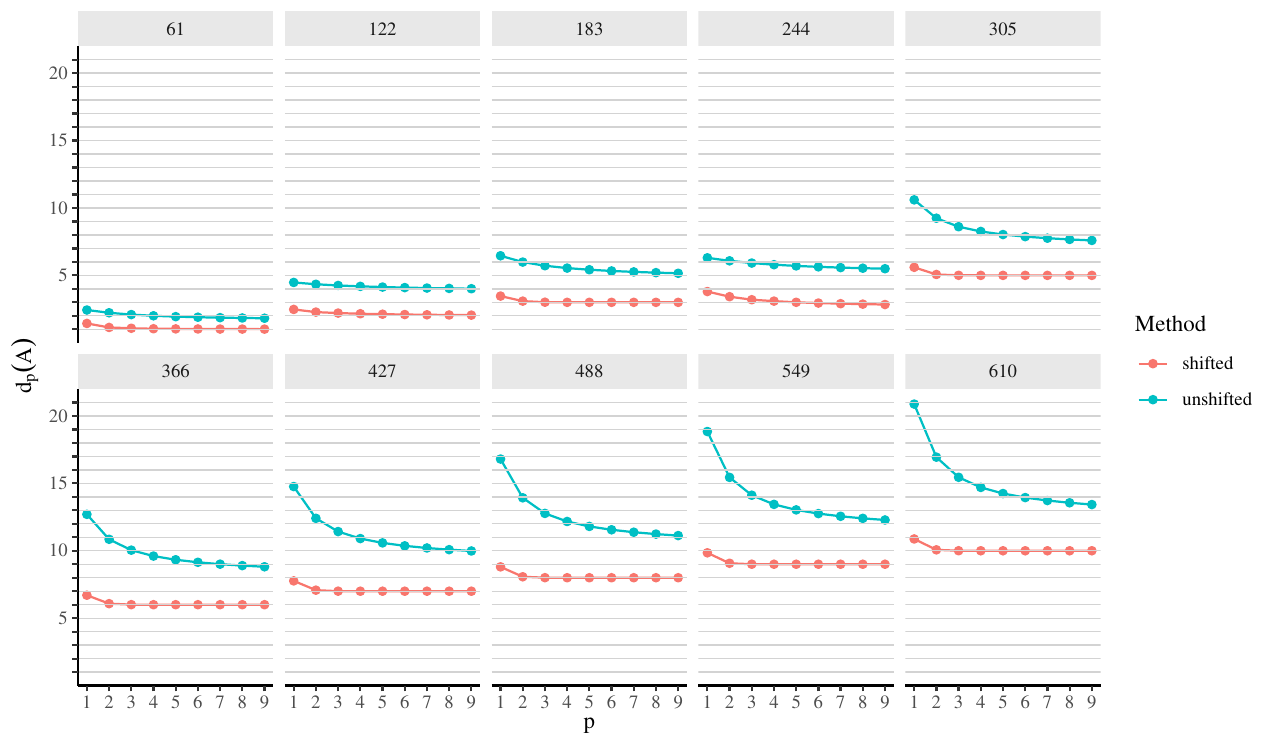}
  \par\end{centering}
  \caption{\label{dvalues-modulated}Estimated $d_p(\matA)=||\matA^p||^{1/p}$ values for Markov-modulated codon models with $\numModules=1,\ldots,10$, with and without shifting.
  Values are averaged over $\matA$ matrices encountered in a short MCMC analysis.
  Shifting reduces the $d_p$ values for these models and makes them nearly flat for $p\ge2$.}
\end{figure}

For the Markov-modulated codon models the $d_p$ values behave differently (Figure~\ref{dvalues-modulated}).
With $\numStates=61\cdot \numModules$, the unshifted $d_p$ values increase with $\numModules$.
The unshifted $d_1$ increases faster than $d_p$.
Thus $d_p$ is fairly flat in $p$ for $\numModules=1$, but $d_p$ has a substantial peak at $p=1$ for $\numModules=10$.
Shifting the rate matrix makes a substantial difference for the Markov-modulated codon models.
The shifted $d_p$ values are lower than the unshifted values.
Additionally, the shifted $d_p$ values are nearly flat so that $d_p\approx \numModules$ for all $p\ge2$.

\subsection{Parallelization strategy}
\beagle splits the work of computing partial likelihoods for each branch among $\numThreads$ threads by dividing sites into $\numThreads$ blocks and assigning each block of sites to a thread.
\beagle also creates a fixed pool of $\numThreads$ worker threads at program initialization and schedules branch-level work on those threads.
We fit the action-based likelihood computation into this framework by dividing columns of the dense matrix $\matB$ among threads.
Performing the multiplication $\matA \cdot \matB$ takes time proportional to the number of columns in $\matB$, so performing $\numThreads$ multiplications with smaller $\matB$ matrices works well.

The speedup from parallelization is limited by the amount of time that is spent running non-parallelized code.
We identified two non-parallelized steps that can take noticeable computation time.
First, the \citet{al2011computing} algorithm approximates $L^1$ norms to determine the values of $m$ and $s$ for each value of $\matQ$.
The computation of these $L^1$ norms is not currently parallelized, so
this serial code provides an upper limit on the amount of parallel speedup that can be achieved using the current strategy.
As the total number of columns $L$ grows, the relative amount of time spent in the parallelized code becomes larger, and the effect of this serial code becomes smaller.

Second, construction of the rate matrix $\matQ$ is not parallelized.
We found that constructing $\matQ$ in dense form, and passing it to \beagle in that form can substantially slow down the action-based algorithm when $\numStates$ is large.
In order to solve this problem, we considered how sparse matrices are stored for computation inside the \beagle library.
The sparse matrix $\matA$ is represented inside \beagle using compressed sparse row (CSR) format.
This format represents the non-zero elements of $\matA$ as a dense array, and uses two corresponding integer arrays to indicate the row and column indices for each non-zero element.
We therefore modify \beast \citep{baele2025beast} to construct the sparse matrix in CSR format and send the matrix to \beagle \citep{ayres2019beagle} in that format.
This implementation substantially decreased the amount of time spent in constructing, storing, and transmitting the $\matQ$ matrix.

\bigskip
\section{Results}
\label{sec:Results}
In order to assess the performance of the action-based likelihood algorithm, we apply it in two scenarios with large numbers of states.
In the first scenario, we examine Markov-modulated codon models \citep{guindon2004modeling, baele2021markov}.
In such models, the substitution process can switch between $\numModules$ different substitution processes, each with $61$ states, leading to $\numStates=\numModules\cdot 61$ states total.
We examine models with $\numModules=1,2,4,8,16,32$.
We are thus able to explore how increasing numbers of states affect the computation time with up to $1\,952$ states.

In the second scenario we consider two paralogs undergoing IGC by jointly modeling the two corresponding codons at each position \citep{ji2016phylogenetic, yang2023interlocus}.
Each position is thus represented by a state space of $61^2 = 3\,721$ states.
The rate matrix for this model has a lower density of nonzero entries than the rate matrices for the Markov-modulated models.

In each scenario, we investigate the parallel efficiency of the action-based likelihood compared to the exponentiation-based likelihood algorithm for $1$ to $10$ threads.
The computation time calculations are based on experiments where we do short MCMC analyses.
The MCMC trajectories are not altered by swapping the likelihood calculation method, and so each run performs the same number of likelihood calculations.

The exponentiation-based likelihood algorithm needs to spend a great deal of time performing $O(S^3)$ matrix multiplications.
\beagle currently uses simple nested for-loops to perform these matrix multiplications for CPUs.
When $\numStates$ is large, we found that these for-loops do not achieve peak efficiency.
We therefore modified the \beagle code to use the C++ matrix library Eigen for matrix multiplication and modified \beast to use Eigen for eigendecomposition. %
These modifications lead to substantial improvements in speed, so
we use these modified versions of \beagle and \beast when measuring the speed of the exponentiation-based algorithm.
For certain tests we also show results from the unmodified version of \beagle, leading to a three-way comparison.

\subsection{Markov Modulated Models}
For Markov-modulated models we analyze a $6$-taxon subset of the yeast data set used in the benchmark of \cite{gangavarapu2024many}.
This data set consists of 931\,844 codon sites that can be classified into 3\,052 site patterns.
To examine the effect of the number of site patterns, we truncated the alignment to obtain data sets with different numbers of unique codon site patterns.
For each value of $\numModules$ and each number of site patterns, we ran the MCMC chain for at least 5 iterations and for at least 5 seconds, and we report the median time per iteration.
This was necessary to achieve stable and consistent numbers despite initial CPU time spent in java startup, including JIT compilation.
We compare the action-based likelihood (`action') to the optimized exponentiation-based implementation (`exp/opt') that uses Eigen for dense matrix operations.

\begin{figure}

  \begin{centering}

  \includegraphics[width=\textwidth]{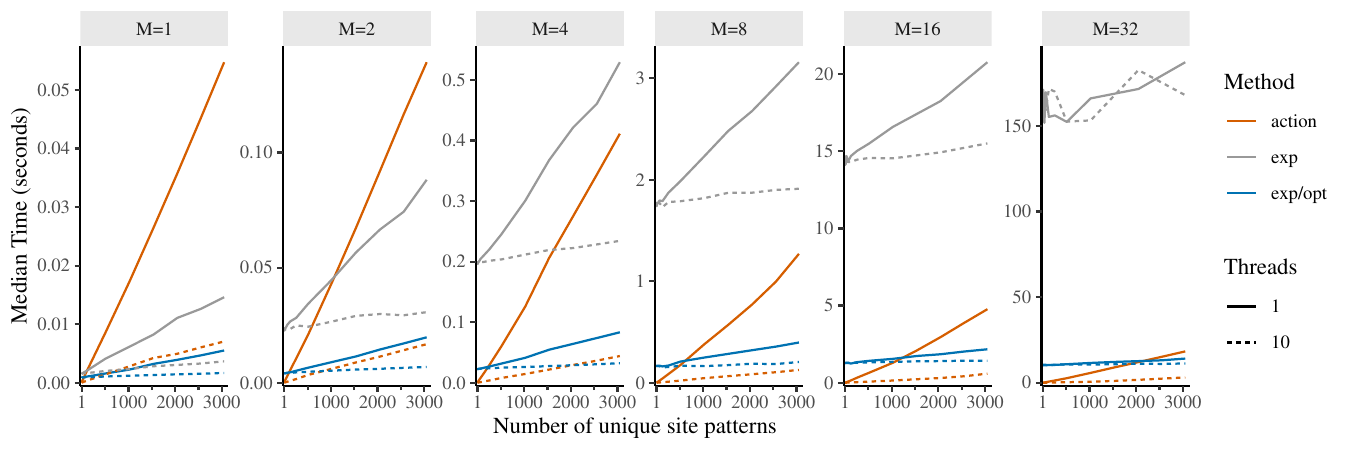}
  \par\end{centering}
\caption{\label{timeVsLength}Computation time versus the number of unique site patterns for Markov-modulated codon models with $\numModules=1,2,4,8,16,32$.
The exp/opt method directly forms the matrix exponential using optimized dense matrix multiplication.}
\end{figure}

\begin{figure}

  \begin{centering}

  \includegraphics[width=\textwidth]{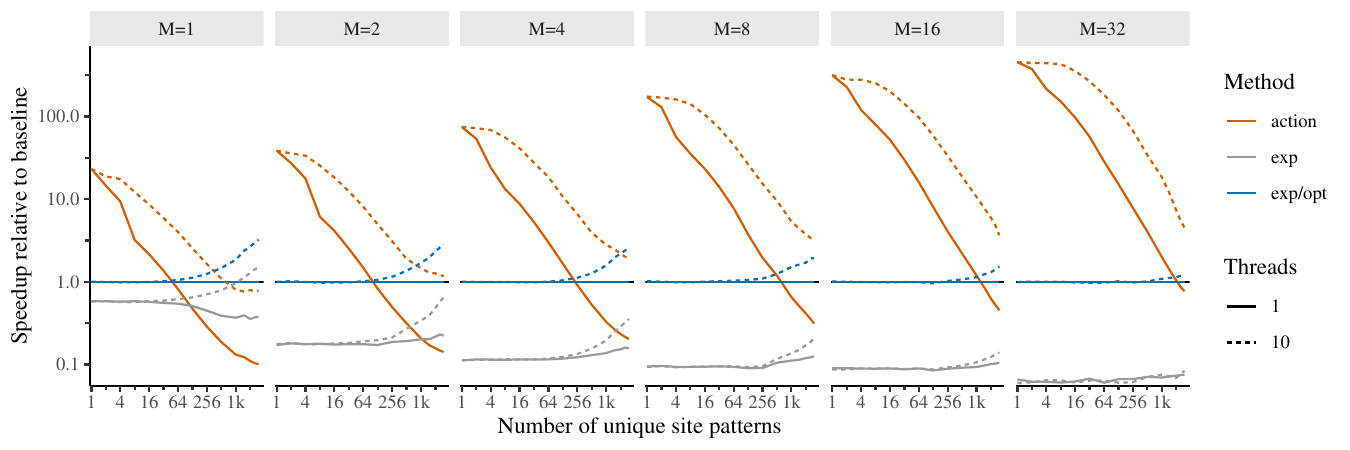}
  \par\end{centering}
\caption{\label{speedupVsLength}Speedup versus the number of unique site patterns for Markov-modulated codon models.  The baseline directly forms the matrix exponential using the optimized dense-matrix implementation and one thread.}

\end{figure}

The exponentiation-based method (`exp/opt') has a substantial fixed cost because it forms dense
transition matrices for each branch, even when the number of site patterns is very small.
The action-based method avoids this cost, but its sparse matrix--dense matrix products scale
directly with the number of site-pattern columns being propagated.
Consequently, it gives its largest speedups when the number of site patterns is small
(Figures~\ref{timeVsLength} and~\ref{speedupVsLength}).
With one thread and a single site pattern, the action-based method is faster than optimized
exponentiation for every model tested, with speedups increasing from about 23-fold for the smallest
model to 454-fold for the largest (Table~\ref{tab:times}).

As the number of site patterns increases, the fixed cost of forming dense transition matrices is
amortized over more columns.
When both methods use one thread, the action-based method has a higher marginal cost per additional
pattern than optimized exponentiation for every model tested.
This additional cost eventually outweighs the saving from avoiding transition-matrix construction,
so the advantage at short alignments disappears.
Larger models retain this advantage over a wider range of pattern counts, so the single-thread
crossover moves to longer alignments as the state space grows.
Nevertheless, on the full 3\,052-pattern data set, optimized exponentiation is faster for every
model when both methods use one thread.

With 10 threads per method, however, the action-based method is faster on the full data set for 
the larger models.
Threading reduces the marginal cost of propagating additional patterns for both methods, but
accelerates a larger fraction of the action-based method's total computation.
Once enough patterns are available to distribute the work across threads, propagation accounts
for most of the action-based method's runtime.
In contrast, the exponentiation-based method retains a substantial fixed cost of eigendecomposition
and transition-matrix construction that remains largely unchanged with thread count in our
implementation.
Threading therefore shifts the crossover to longer alignments: the action-based method can have a
lower total runtime despite retaining a higher marginal cost per pattern at equal thread counts.
For the largest model on the full data set, the action-based method changes from about 1.3 times slower
with one thread per method to 3.6 times faster with 10 threads per method
(Figure~\ref{timeVsLength}).
These gains require enough patterns to share the work; for a single pattern, additional threads
provide little benefit.

As a supporting fixed-length data comparison, we also measured total computation time for the same short MCMC analysis on the full 3\,052-pattern data set, rather than time per iteration.
Those fixed-length run timings show the same qualitative pattern: the cost of direct exponentiation grows rapidly with the number of states, while the action-based method benefits more from threading because most of its work is in site-pattern-parallel matrix products.
The fixed-length run figures are shown in the Appendix.

\subsection{Two-paralog IGC model}
For the two-paralog IGC numerical example we employ the same data set of yeast paralogous ribosomal protein coding genes YLR406C and YDL075W that are studied in \cite{ji2016phylogenetic} and \cite{yang2023interlocus}.
The two-paralog IGC model has $3\,721$ states.
The alignment consists of 112 codon sites, which comprise only 90 unique site patterns.
To examine the effect of the number of site patterns, we generated IGC benchmark data sets with different numbers of site-pattern columns and report the median time per MCMC iteration.
In order to obtain computation times $>90$ unique site patterns, we disable site compression and cyclically repeat the observed sites until we reach the desired length.
As in the Markov-modulated analysis, we compare the action-based likelihood to the optimized exponentiation-based implementation that uses Eigen for dense matrix operations.
We do not include the unoptimized exponentiation-based implementation in the IGC length benchmarks because its runtime for the 3\,721-state model was prohibitively large.

\begin{figure}
  \begin{centering}
  \includegraphics[width=0.48\textwidth]{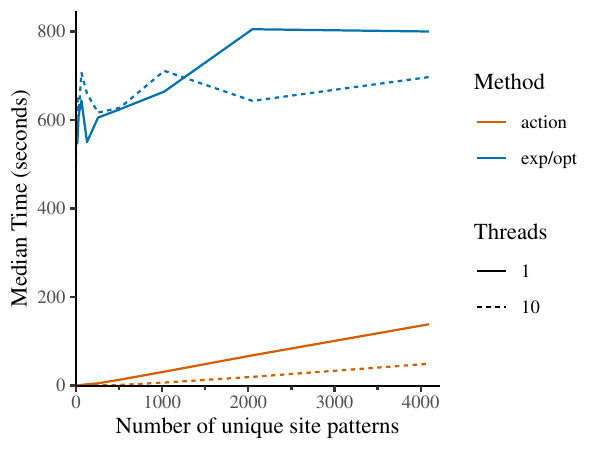}
  \includegraphics[width=0.48\textwidth]{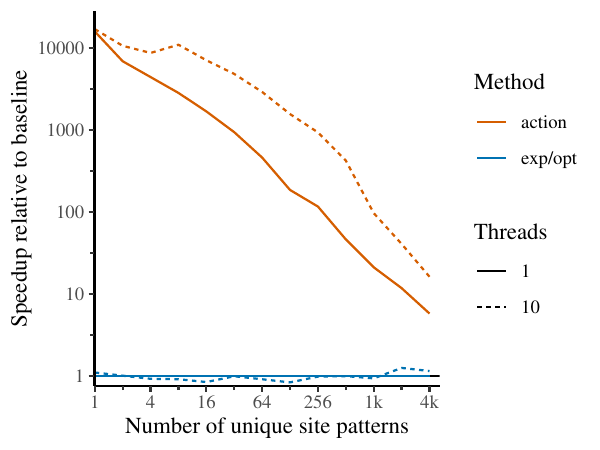}
  \par\end{centering}
\caption{\label{fig:IGClength}Computation time and speedup versus the number of unique site patterns for the IGC two-paralog data set.
Left: computation time.
Right: speedup relative to the optimized one-thread exponentiation baseline.}
\end{figure}

The IGC model shows a much larger advantage for the action-based method than the Markov-modulated codon models (Figure~\ref{fig:IGClength}).
The reason is that the IGC state space is much larger and the rate matrix is substantially sparser.
Direct exponentiation therefore pays a very large cost to form dense transition matrices, even when only a small number of site patterns is propagated.
The action-based method avoids this fixed dense-matrix cost and instead spends most of its time in sparse matrix--dense matrix products.

As the number of site patterns increases, the action-based runtime increases because more columns must be propagated through the Taylor approximation.
The speedup therefore decreases with the number of site patterns, as in the Markov-modulated example.
However, the speedup remains large over the plotted range because the IGC rate matrix is sufficiently sparse that direct exponentiation remains dominated by the cost of dense matrix operations.
In this case the advantage is not only a consequence of avoiding the fixed cost of forming transition matrices.
Across the benchmarked range, the action-based method also appears to have a lower marginal cost per additional site pattern than optimized exponentiation.
The relative speedup therefore decreases as the fixed cost of exponentiation is amortized, but optimized exponentiation remains substantially slower throughout the plotted range.
Threading again has little effect when the number of site patterns is very small, but substantially improves the action-based runtime once there are enough columns to divide across threads.

On the original 90-pattern data set, a fixed-run benchmark that reports total runtime for the same short MCMC analysis shows that the exponentiation-based likelihood is about 889 times slower than the action-based likelihood with 1 thread, and 3\,623 times slower with 10 threads.
The exponentiation-based likelihood only achieves a speedup of 1.72 times by using 10 threads, because much of its time is spent in matrix-exponential calculations that are not parallelized over site patterns.
In contrast, the action-based likelihood achieves a 5.86-fold speedup by using 10 threads.
This speedup is less than 10-fold because time spent estimating the L1 norms $d_p(\matA)$ is not currently parallelized.
The L1-norm computations take a relatively larger fraction of the computation time in this example than in the Markov-modulated example because the original IGC data set has far fewer site patterns.
The corresponding fixed-run threading figure is shown in the Appendix.

Representative speedups are summarized in Table~\ref{tab:times}.
These rows show how the advantage of the action-based method depends jointly on the state space size, rate-matrix density, and number of site patterns.

\begin{table}[H]
	\centering
	\setlength{\tabcolsep}{4pt}
\begin{tabular}{lrrrrrrS[table-format=1.2]}
	\toprule
	Model & $\numStates$ & $\#>0$ & density & patterns & \multicolumn{2}{c}{Baseline speedup} & \multicolumn{1}{c}{Thread speedup} \\
	\cmidrule(lr){6-7}\cmidrule(l){8-8}
	& & & & & \multicolumn{1}{r}{1 Thr} & \multicolumn{1}{r}{10 Thr} & \multicolumn{1}{c}{10 vs 1 Thr} \\
	\midrule
	MM, $\numModules=1$ & 61 & 587 & 15.78\% & 1 & 23.2 & 23.2 & 1.00 \\
	MM, $\numModules=1$ & 61 & 587 & 15.78\% & 64 & 0.8 & 4.0 & 4.91 \\
	MM, $\numModules=1$ & 61 & 587 & 15.78\% & 512 & 0.2 & 1.1 & 5.87 \\
	MM, $\numModules=1$ & 61 & 587 & 15.78\% & 3\,052 & 0.1 & 0.8 & 7.73 \\
	MM, $\numModules=8$ & 488 & 8\,112 & 3.41\% & 1 & 172.4 & 174.5 & 1.01 \\
	MM, $\numModules=8$ & 488 & 8\,112 & 3.41\% & 64 & 7.8 & 45.7 & 5.85 \\
	MM, $\numModules=8$ & 488 & 8\,112 & 3.41\% & 512 & 1.2 & 9.6 & 8.02 \\
	MM, $\numModules=8$ & 488 & 8\,112 & 3.41\% & 3\,052 & 0.3 & 3.1 & 9.97 \\
	MM, $\numModules=32$ & 1\,952 & 79\,296 & 2.08\% & 1 & 454.1 & 454.3 & 1.00 \\
	MM, $\numModules=32$ & 1\,952 & 79\,296 & 2.08\% & 64 & 28.7 & 183.8 & 6.40 \\
	MM, $\numModules=32$ & 1\,952 & 79\,296 & 2.08\% & 512 & 4.1 & 34.2 & 8.41 \\
	MM, $\numModules=32$ & 1\,952 & 79\,296 & 2.08\% & 3\,052 & 0.8 & 4.5 & 5.87 \\
	IGC & 3\,721 & 74\,161 & 0.54\% & 1 & 15\,960.0 & 17\,064.7 & 1.07 \\
	IGC & 3\,721 & 74\,161 & 0.54\% & 64 & 462.1 & 2\,916.6 & 6.31 \\
	IGC & 3\,721 & 74\,161 & 0.54\% & 512 & 46.4 & 424.7 & 9.16 \\
	IGC & 3\,721 & 74\,161 & 0.54\% & 4\,096 & 5.8 & 16.2 & 2.81 \\
	\bottomrule
\end{tabular}

	\caption{
	Representative speedups for action-based likelihood computation.
	Speedups are computed as the time for optimized exponentiation with one thread divided by the time for the action-based method with the indicated number of threads, matched by model and number of site patterns.
	The final column reports the action-based runtime with one thread divided by the action-based runtime with 10 threads; the ideal value is 10.
	All rows are from the time-per-iteration length benchmarks.
	\label{tab:times}
}
\end{table}

\bigskip

\section{Discussion}
\label{sec4}
We introduce a new method for computing phylogenetic likelihoods that can be substantially faster
for models with large, sparse state spaces, particularly when the number of site patterns is small.
The action-based likelihood method is based on computing $e^{\matQ \cdot b_i}\cdot\mathbf{p}$ without ever directly forming the transition probability matrix $e^{\matQ \cdot b_i}$.
We implement this approach for CPUs in the \beagle library.
By distributing pattern propagation across CPU cores, our implementation accelerates most of the
action-based computation when enough patterns are present, extending its advantage to longer
alignments even when both methods use the same number of threads.
The action-based method achieves speedups of up to 17\,000-fold compared to the traditional exponentiation-based approach on models with 3\,721 states.

Computing the action of the matrix exponential offers the largest speed increases when $\matQ$ is sparse.
The algorithm spends most of its computation time performing multiplications of the form $\matA \cdot \matB$, where $\matA=\matQ\cdot b_i$ is sparse and $\matB$ is dense.
When only a fraction $\density$ of the entries in $\matA$ are non-zero, the cost of this multiplication is $O(\density\cdot \numStates^2 \cdot \numSites)$.
However, the degree of sparsity depends on the model, so not all models benefit equally from the action-based approach.
For example, the density of nonzero entries in the IGC two-paralog model is about 4 times lower than in the Markov-modulated codon model with $\numModules=32$, so the Markov-modulated models benefit less from the action-based algorithm than the IGC model.

\subsection{Efficient dense matrix multiplication}
Traditional phylogenetic likelihood methods that form the matrix exponential directly perform a large number of matrix multiplications.
These libraries usually compute an eigendecomposition $(\matV,\matD)$ of $\matQ$ where $\matQ = \matV\cdot \matD\cdot \matV^{-1}$.
When $\matQ$ has complex eigenvalues, the matrix exponentiation operation needs special treatment regarding the conjugate complex eigenvalue pairs \citep[see e.g., the supplementary material of~][ for more detailed discussions]{ji2025detecting}.
Additionally, some rate matrices are not diagonalizable at all. %
This issue does not arise for the matrix action calculations because eigendecomposition is not involved.

Once the eigendecomposition is computed, the transition probability matrix for a branch can be computed as $\matV\cdot e^{\matD\cdot  b_i}\cdot \matV^{-1}$.
This calculation involves a matrix multiplication that has a cost of $O(\numStates^3)$, and must be performed whenever a branch length $b_i$ changes or $\matQ$ changes.
Computing the eigendecomposition also has an $O(\numStates^3)$ cost, but only needs to be performed when $\matQ$ changes.
The speed of traditional phylogenetic likelihood methods thus depends strongly on the efficiency of the implementation of dense matrix multiplication when the number of states is large.

Both \beagle and the Phylogenetic Likelihood Library (PLL) \citep{flouri2015phylogenetic} implement matrix multiplication using simple nested for-loops.
PLL additionally implements SSE and AVX versions of these loops.
Optimized implementations of GEMM, the general matrix-multiplication operation in Level 3 BLAS
\citep{dongarra1990set}, divide matrices into blocks to reuse entries while they remain in cache
\citep{goto2008anatomy}.
Matrix multiplication accesses $O(\numStates^2)$ distinct matrix entries but performs
$O(\numStates^3)$ arithmetic operations.
Without blocking, entries may be fetched from memory repeatedly, limiting performance by memory
access rather than arithmetic speed.
We therefore implement efficient matrix multiplication using the Eigen C++ library for matrices \citep{eigenweb}.
This allows us to perform a fair comparison between our action-based method and methods that directly form the matrix exponential.

Equivalently, if $\matB$ is a matrix whose columns are partial-likelihood vectors, one can compute propagated partial likelihoods as $\matB^\prime=\matV\cdot(e^{\matD\cdot  b_i}\cdot(\matV^{-1}\cdot \matB))$.
Although this approach relies on an eigendecomposition, it does not actually perform the $O(\numStates^3)$ computation to compute the exponential matrix.
For short sequences, it is therefore faster than computing the matrix exponential.
However, when $\numSites>\approx\numStates$ for some constant $\gamma$, computing the exponential matrix becomes faster.
This method does not take advantage of sparsity in the rate matrix $\matQ$.

\subsection{Effects of sequence length, rate-matrix structure, and threading}
The sequence length benchmarks show that the balance between fixed and marginal costs depends on
threading as well as the model and data.
With one thread, the higher marginal cost of the action-based method eventually outweighs the
saving from avoiding transition-matrix construction for every Markov-modulated model tested.
With multiple threads, however, a larger fraction of its total runtime is accelerated, while the
exponentiation-based implementation retains a substantial fixed cost.
Consequently, the action-based method can remain faster at longer alignments even though its
marginal cost is still higher at equal thread counts, as observed for the larger models.
The action-based method also retains serial work, including rate-matrix construction and norm
estimation.
These results describe the current implementations; parallelizing eigendecomposition or
transition-matrix construction could change the comparison.
For the much sparser IGC model, however, the action-based method appears to have both a lower fixed
cost and a lower marginal cost per pattern, and remains substantially faster throughout the
measured range.
Which method is faster therefore depends on the rate-matrix structure, the number of unique site
patterns and states, and the number of threads.

\subsection{Effects of branch length}
The present benchmarks do not systematically examine how branch length affects the cost of the action-based likelihood.
This is an important direction for future work because MCMC algorithms sometimes propose very long branch lengths, and the likelihood must still be computed accurately in order to reject such proposals.
For a branch of length $b_i$, the action algorithm is applied to $\matA=\matQ b_i$ (or to $\matQ\gamma_r b_i$ for a rate class), so the relevant matrix norms scale with branch length.
Longer branches can therefore require a larger Taylor degree $m$ or, once $m$ is near its useful maximum, a larger scaling factor $s$.
Since the runtime of the action algorithm is proportional to roughly $s\cdot m$ sparse matrix--dense matrix multiplications per branch, the cost may grow approximately linearly with branch length in regimes where $s$ dominates.
Quantifying this behavior, and determining whether long-branch proposals need special handling, remains future work.

\subsection{Choosing and reusing Taylor parameters}
The method by \citet{al2011computing} attempts to determine whether or not the amount of time saved using $d_p(\matA)$ exceeds the amount of time lost by computing the $d_p(\matA)$ values.
This is done using code fragment 3.13 in the original paper.
The calculation depends on the number of columns in the data set.
However, it does not take into account that the $d_p(\matA)$ can be re-used several times.
Nor does it take into account the fact that computation of the $d_p(\matA)$ is single-threaded, while the computation of $e^{\matA}\cdot \matB$ can be multi-threaded.
Lastly, since computing $d_p(\matA)$ involves computing $\matA^p\cdot \mathbf{X}$ for a vector X, one could possibly share work between the computation of $d_p(\matA)$ and $d_{p+1}(\matA)$.
Code fragment 3.13 does not account for this either.
We therefore cache $d_p(\matA)$ values and re-compute them only when $\matA$ changes.
However, we leave other adjustments (e.g.,~parallelization of $d_p(\matA)$ computation) for future work.

\subsection{Broader implications}
More broadly, large sparse state space models remain underused in evolutionary biology because their computational costs have made them difficult to fit routinely.
Action-based likelihood computation reduces one of the main barriers to using these models in molecular evolution, biogeography, and other settings where richer state spaces can better represent the biological process.

\bigskip
\section{Materials and Methods}
\label{sec:MatMeth}

To assess the performance of the action-based and the traditional likelihood computation, we ran MCMC computations in \beast \citep{baele2025beast} with each likelihood algorithm.
Run-times were computed on a lightly loaded AMD Ryzen 9 7900X 12-Core Processor. %

For each benchmark comparison, the data, starting values, proposal schedule, and model settings were held fixed while changing the likelihood implementation and the number of CPU threads.
The length benchmarks varied the number of site-pattern columns and report median wall-clock time per MCMC iteration.
The benchmarks in the Appendix did not vary the length of the data sets; they used the original or full benchmark data sets and report total wall-clock time for the same short MCMC run. 
The timing data and plotting scripts used to generate the benchmark figures are included in the repository at \url{https://github.com/ji-group/ActionCPUSupplementary/} under \texttt{data/} and \texttt{scripts/}.

Running \beast computations with the action-based likelihood requires using the \texttt{action\_likelihood} branch of beast from \url{https://github.com/beast-dev/beast-mcmc.git} and the \texttt{action-dev} branch of \beagle available from \url{https://github.com/ji-group/beagle-lib/}.

\beast computations did not originally use either the Eigen library or a BLAS implementation.
We implemented a JNI library so that eigendecompositions could be done with Eigen.
This JNI library is available at \url{https://github.com/ji-group/EigenJNI}.
We additionally modified \beagle to improve the speed of dense matrix multiplication.
These changes are on branch \texttt{calc-partials-eigen}, which includes the changes from branch \texttt{action-dev} from \url{git@github.com:ji-group/beagle-lib.git}.

To assess how much computation time was spent in specific functions, we used the Linux \texttt{perf} subsystem.
We used the \texttt{hotspot} software to visualize the results and identify non-parallelized regions of code, and what task they were performing.

\section{Acknowledgements}

MAS and XJ are partially supported by NIH grants U19 R01 AI135995, AI153044 and R01 AI162611.
XJ acknowledges support from the RCS program of Louisiana Board of Regents grant, NSF DEB1754142 and NIH R01GM072562.
GB and PL acknowledge support from the Internal Funds KU Leuven (Grant No. C16/26/012).
GB acknowledges support from the Research Foundation - Flanders (``Fonds voor Wetenschappelijk Onderzoek - Vlaanderen'', G098321N), from the European Union Horizon 2023 RIA project LEAPS (grant agreement no. 101094685) and from the DURABLE EU4Health project 02/2023-01/2027, which is co-funded by the European Union (call EU4H-2021-PJ4) under Grant Agreement No. 101102733.
PL acknowledges support by the Research Foundation -- Flanders (``Fonds voor Wetenschappelijk Onderzoek -- Vlaanderen'', G010326N and G051322N).
KG is partially supported by US National Institutes of Health (NIH) grant U19 AI135995.
We gratefully acknowledge support from NVIDIA Corporation and Advanced Micro Devices, Inc.~with the donation of parallel computing resources used for this research.

\bigskip\bigskip

\bibliographystyle{sysbio}
\bibliography{action}

\begin{thebibliography}{42}
\providecommand{\natexlab}[1]{#1}
\providecommand{\selectlanguage}[1]{\relax}
\providecommand{\bibAnnoteFile}[1]{%
  \IfFileExists{#1}{\begin{quotation}\noindent\textsc{Key:} #1\\
  \textsc{Annotation:}\ \input{#1}\end{quotation}}{}}
\providecommand{\bibAnnote}[2]{%
  \begin{quotation}\noindent\textsc{Key:} #1\\
  \textsc{Annotation:}\ #2\end{quotation}}

\bibitem[{Al-Mohy and Higham(2011)}]{al2011computing}
Al-Mohy, A.~H. and N.~J. Higham. 2011. Computing the action of the matrix
  exponential, with an application to exponential integrators. SIAM journal on
  scientific computing 33:488–511.
\bibAnnoteFile{al2011computing}

\bibitem[{Ayres et~al.(2012)Ayres, Darling, Zwickl, Beerli, Holder, Lewis,
  Huelsenbeck, Ronquist, Swofford, Cummings et~al.}]{ayres2012beagle}
Ayres, D., A.~Darling, D.~Zwickl, P.~Beerli, M.~Holder, P.~Lewis,
  J.~Huelsenbeck, F.~Ronquist, D.~Swofford, M.~Cummings, et~al. 2012. {BEAGLE}:
  an application programming interface and high-performance computing library
  for statistical phylogenetics. Systematic Biology 61:170–173 {PMCID}:
  PMC3243739.
\bibAnnoteFile{ayres2012beagle}

\bibitem[{Ayres et~al.(2019)Ayres, Cummings, Baele, Darling, Lewis, Swofford,
  Huelsenbeck, Lemey, Rambaut, and Suchard}]{ayres2019beagle}
Ayres, D.~L., M.~P. Cummings, G.~Baele, A.~E. Darling, P.~O. Lewis, D.~L.
  Swofford, J.~P. Huelsenbeck, P.~Lemey, A.~Rambaut, and M.~A. Suchard. 2019.
  Beagle 3: improved performance, scaling, and usability for a high-performance
  computing library for statistical phylogenetics. Systematic biology
  68:1052–1061.
\bibAnnoteFile{ayres2019beagle}

\bibitem[{Baele et~al.(2021)Baele, Gill, Bastide, Lemey, and
  Suchard}]{baele2021markov}
Baele, G., M.~S. Gill, P.~Bastide, P.~Lemey, and M.~A. Suchard. 2021.
  Markov-modulated continuous-time markov chains to identify site-and
  branch-specific evolutionary variation in beast. Systematic Biology
  70:181–189.
\bibAnnoteFile{baele2021markov}

\bibitem[{Baele et~al.(2025)Baele, Ji, Hassler, McCrone, Shao, Zhang, Holbrook,
  Lemey, Drummond, Rambaut et~al.}]{baele2025beast}
Baele, G., X.~Ji, G.~W. Hassler, J.~T. McCrone, Y.~Shao, Z.~Zhang, A.~J.
  Holbrook, P.~Lemey, A.~J. Drummond, A.~Rambaut, et~al. 2025. Beast x for
  bayesian phylogenetic, phylogeographic and phylodynamic inference. Nature
  Methods 22:1653–1656.
\bibAnnoteFile{baele2025beast}

\bibitem[{Bettisworth et~al.(2023)Bettisworth, Smith, and
  Stamatakis}]{bettisworth2023lagrange}
Bettisworth, B., S.~A. Smith, and A.~Stamatakis. 2023. Lagrange-ng: The next
  generation of lagrange. Systematic Biology 72:242--248.
\bibAnnoteFile{bettisworth2023lagrange}

\bibitem[{Bi et~al.(2026)Bi, Li, Qiu, Xu, Yang, Yuan, Dong, Xiao, and
  Liu}]{bi2026memoryaware}
Bi, D., S.~Li, H.~Qiu, C.~Xu, X.~Yang, F.~Yuan, D.~Dong, T.~Xiao, and J.~Liu.
  2026. A memory-aware sparse matrix-matrix multiplication on multicore
  architectures. ACM Transactions on Architecture and Code Optimization 23.
\bibAnnoteFile{bi2026memoryaware}

\bibitem[{Borges et~al.(2022)Borges, Boussau, Szöllősi, and
  Kosiol}]{borges2022nucleotide}
Borges, R., B.~Boussau, G.~J. Szöllősi, and C.~Kosiol. 2022. Nucleotide usage
  biases distort inferences of the species tree. Genome biology and evolution
  14:evab290.
\bibAnnoteFile{borges2022nucleotide}

\bibitem[{Boucher and D{\'e}mery(2016)}]{boucher2016inferring}
Boucher, F.~C. and V.~D{\'e}mery. 2016. Inferring bounded evolution in
  phenotypic characters from phylogenetic comparative data. Systematic biology
  65:651--661.
\bibAnnoteFile{boucher2016inferring}

\bibitem[{Boucher et~al.(2018)Boucher, D{\'e}mery, Conti, Harmon, and
  Uyeda}]{boucher2018general}
Boucher, F.~C., V.~D{\'e}mery, E.~Conti, L.~J. Harmon, and J.~Uyeda. 2018. A
  general model for estimating macroevolutionary landscapes. Systematic biology
  67:304--319.
\bibAnnoteFile{boucher2018general}

\bibitem[{Calabrese and Sainudiin(2005)}]{calabrese2005models}
Calabrese, P. and R.~Sainudiin. 2005. Models of microsatellite evolution.
  Pages~290--305 \emph{in} Statistical methods in molecular evolution.
  Springer.
\bibAnnoteFile{calabrese2005models}

\bibitem[{Davydov et~al.(2017)Davydov, Robinson-Rechavi, and
  Salamin}]{davydov2017state}
Davydov, I.~I., M.~Robinson-Rechavi, and N.~Salamin. 2017. State aggregation
  for fast likelihood computations in molecular evolution. Bioinformatics
  33:354--362.
\bibAnnoteFile{davydov2017state}

\bibitem[{De~Bie et~al.(2006)De~Bie, Cristianini, Demuth, and
  Hahn}]{deBie2006cafe}
De~Bie, T., N.~Cristianini, J.~P. Demuth, and M.~W. Hahn. 2006. Cafe: a
  computational tool for the study of gene family evolution. Bioinformatics
  22:1269--1271.
\bibAnnoteFile{deBie2006cafe}

\bibitem[{de~Koning et~al.(2010)de~Koning, Gu, and Pollock}]{deKoning2010rapid}
de~Koning, A.~J., W.~Gu, and D.~D. Pollock. 2010. Rapid likelihood analysis on
  large phylogenies using partial sampling of substitution histories. Molecular
  biology and evolution 27:249--265.
\bibAnnoteFile{deKoning2010rapid}

\bibitem[{{De Maio} et~al.(2015){De Maio}, Schrempf, and Kosiol}]{de2015pomo}
{De Maio}, N., D.~Schrempf, and C.~Kosiol. 2015. Pomo: an allele
  frequency-based approach for species tree estimation. Systematic biology
  64:1018–1031.
\bibAnnoteFile{de2015pomo}

\bibitem[{Dib et~al.(2014)Dib, Silvestro, and Salamin}]{dib2014evolutionary}
Dib, L., D.~Silvestro, and N.~Salamin. 2014. Evolutionary footprint of
  coevolving positions in genes. Bioinformatics 30:1241–1249.
\bibAnnoteFile{dib2014evolutionary}

\bibitem[{Dongarra et~al.(1990)Dongarra, Du~Croz, Hammarling, and
  Duff}]{dongarra1990set}
Dongarra, J.~J., J.~Du~Croz, S.~Hammarling, and I.~S. Duff. 1990. A set of
  level 3 basic linear algebra subprograms. ACM Transactions on Mathematical
  Software 16:1--17.
\bibAnnoteFile{dongarra1990set}

\bibitem[{Felsenstein(1981)}]{felsenstein1981evolutionary}
Felsenstein, J. 1981. Evolutionary trees from {DNA} sequences: a maximum
  likelihood approach. Journal of Molecular Evolution 17:368–376.
\bibAnnoteFile{felsenstein1981evolutionary}

\bibitem[{Flouri et~al.(2015)Flouri, Izquierdo-Carrasco, Darriba, Aberer,
  Nguyen, Minh, von Haeseler, and Stamatakis}]{flouri2015phylogenetic}
Flouri, T., F.~Izquierdo-Carrasco, D.~Darriba, A.~J. Aberer, L.-T. Nguyen,
  B.~Q. Minh, A.~von Haeseler, and A.~Stamatakis. 2015. The phylogenetic
  likelihood library. Systematic Biology 64:356–362.
\bibAnnoteFile{flouri2015phylogenetic}

\bibitem[{Gangavarapu et~al.(2024)Gangavarapu, Ji, Baele, Fourment, Lemey,
  {Matsen IV}, and Suchard}]{gangavarapu2024many}
Gangavarapu, K., X.~Ji, G.~Baele, M.~Fourment, P.~Lemey, F.~A. {Matsen IV}, and
  M.~A. Suchard. 2024. Many-core algorithms for high-dimensional gradients on
  phylogenetic trees. Bioinformatics 40:btae030.
\bibAnnoteFile{gangavarapu2024many}

\bibitem[{Gangavarapu et~al.(2026)Gangavarapu, Ji, Shao, Lemey, Rambaut, Baele,
  and Suchard}]{gangavarapu2026beagle}
Gangavarapu, K., X.~Ji, Y.~Shao, P.~Lemey, A.~Rambaut, G.~Baele, and M.~A.
  Suchard. 2026. Beagle 4.1: A high-performance library for computation on
  phylogenetic trees across diverse parallel architectures.
\bibAnnoteFile{gangavarapu2026beagle}

\bibitem[{Glick and Mayrose(2014)}]{glick2015chromevol}
Glick, L. and I.~Mayrose. 2014. Chromevol: assessing the pattern of chromosome
  number evolution and the inference of polyploidy along a phylogeny. {Mol Biol
  Evol} 31:1914–22.
\bibAnnoteFile{glick2015chromevol}

\bibitem[{Goto and van~de Geijn(2008)}]{goto2008anatomy}
Goto, K. and R.~A. van~de Geijn. 2008. Anatomy of high-performance matrix
  multiplication. ACM Transactions on Mathematical Software 34:12:1--12:25.
\bibAnnoteFile{goto2008anatomy}

\bibitem[{Guennebaud et~al.(2010)Guennebaud, Jacob et~al.}]{eigenweb}
Guennebaud, G., B.~Jacob, et~al. 2010. Eigen v3. http://eigen.tuxfamily.org.
\bibAnnoteFile{eigenweb}

\bibitem[{Guindon et~al.(2004)Guindon, Rodrigo, Dyer, and
  Huelsenbeck}]{guindon2004modeling}
Guindon, S., A.~G. Rodrigo, K.~A. Dyer, and J.~P. Huelsenbeck. 2004. Modeling
  the site-specific variation of selection patterns along lineages. Proceedings
  of the National Academy of Sciences 101:12957–12962.
\bibAnnoteFile{guindon2004modeling}

\bibitem[{Heddes et~al.(2025)Heddes, Srinivasa, Givargis, and
  Nicolau}]{heddes2025alwayssparse}
Heddes, M., N.~Srinivasa, T.~Givargis, and A.~Nicolau. 2025. Always-sparse
  training by growing connections with guided stochastic exploration. Appendix
  K.
\bibAnnoteFile{heddes2025alwayssparse}

\bibitem[{Higham and Tisseur(2000)}]{higham2000block}
Higham, N.~J. and F.~Tisseur. 2000. A block algorithm for matrix 1-norm
  estimation, with an application to 1-norm pseudospectra. SIAM Journal on
  Matrix Analysis and Applications 21:1185–1201.
\bibAnnoteFile{higham2000block}

\bibitem[{Huelsenbeck(2002)}]{huelsenbeck2002testing}
Huelsenbeck, J.~P. 2002. Testing a covariotide model of dna substitution.
  Molecular biology and evolution 19:698–707.
\bibAnnoteFile{huelsenbeck2002testing}

\bibitem[{Irvahn and Minin(2014)}]{irvahn2014phylogenetic}
Irvahn, J. and V.~N. Minin. 2014. Phylogenetic stochastic mapping without
  matrix exponentiation. Journal of Computational Biology 21:676–690.
\bibAnnoteFile{irvahn2014phylogenetic}

\bibitem[{Ji et~al.(2016)Ji, Griffing, and Thorne}]{ji2016phylogenetic}
Ji, X., A.~Griffing, and J.~L. Thorne. 2016. A phylogenetic approach finds
  abundant interlocus gene conversion in yeast. Molecular biology and evolution
  33:2469–2476.
\bibAnnoteFile{ji2016phylogenetic}

\bibitem[{Ji et~al.(2025)Ji, Redelings, Su, Bao, Deng, Monti, Hong, Baele,
  Lemey, and Suchard}]{ji2025detecting}
Ji, X., B.~Redelings, S.~Su, H.~Bao, W.-M. Deng, F.~Monti, S.~L. Hong,
  G.~Baele, P.~Lemey, and M.~A. Suchard. 2025. Detecting evolutionary
  change-points with branch-specific substitution models and shrinkage priors.
  Preprint, version 2.
\bibAnnoteFile{ji2025detecting}

\bibitem[{Leventhal et~al.(2014)Leventhal, G{\"u}nthard, Bonhoeffer, and
  Stadler}]{leventhal2014using}
Leventhal, G.~E., H.~F. G{\"u}nthard, S.~Bonhoeffer, and T.~Stadler. 2014.
  Using an epidemiological model for phylogenetic inference reveals density
  dependence in hiv transmission. Molecular biology and evolution 31:6--17.
\bibAnnoteFile{leventhal2014using}

\bibitem[{Mateiu and Rannala(2006)}]{mateiu2006inferring}
Mateiu, L. and B.~Rannala. 2006. Inferring complex dna substitution processes
  on phylogenies using uniformization and data augmentation. Systematic biology
  55:259--269.
\bibAnnoteFile{mateiu2006inferring}

\bibitem[{Mayrose et~al.(2010)Mayrose, Barker, and
  Otto}]{mayrose2010probabilistic}
Mayrose, I., M.~S. Barker, and S.~P. Otto. 2010. Probabilistic models of
  chromosome number evolution and the inference of polyploidy. {Syst Biol}
  59:132–44.
\bibAnnoteFile{mayrose2010probabilistic}

\bibitem[{Ree and Smith(2008)}]{ree2008maximum}
Ree, R.~H. and S.~A. Smith. 2008. Maximum likelihood inference of geographic
  range evolution by dispersal, local extinction, and cladogenesis. Systematic
  biology 57:4–14.
\bibAnnoteFile{ree2008maximum}

\bibitem[{Rodrigue et~al.(2009)Rodrigue, Kleinman, Philippe, and
  Lartillot}]{rodrigue2009computational}
Rodrigue, N., C.~L. Kleinman, H.~Philippe, and N.~Lartillot. 2009.
  Computational methods for evaluating phylogenetic models of coding sequence
  evolution with dependence between codons. Molecular Biology and Evolution
  26:1663–1676.
\bibAnnoteFile{rodrigue2009computational}

\bibitem[{Schrempf et~al.(2016)Schrempf, Minh, {De Maio}, von Haeseler, and
  Kosiol}]{schrempf2016reversible}
Schrempf, D., B.~Q. Minh, N.~{De Maio}, A.~von Haeseler, and C.~Kosiol. 2016.
  Reversible polymorphism-aware phylogenetic models and their application to
  tree inference. bioRxiv .
\bibAnnoteFile{schrempf2016reversible}

\bibitem[{Sidje(1998)}]{sidje1998expokit}
Sidje, R.~B. 1998. Expokit: A software package for computing matrix
  exponentials. ACM Transactions on Mathematical Software (TOMS) 24:130--156.
\bibAnnoteFile{sidje1998expokit}

\bibitem[{Tuffley and Steel(1998)}]{tuffley1998modeling}
Tuffley, C. and M.~Steel. 1998. Modeling the covarion hypothesis of nucleotide
  substitution. Mathematical biosciences 147:63–91.
\bibAnnoteFile{tuffley1998modeling}

\bibitem[{Wu and Drummond(2011)}]{wu2011joint}
Wu, C.-H. and A.~J. Drummond. 2011. Joint inference of microsatellite mutation
  models, population history and genealogies using transdimensional markov
  chain monte carlo. Genetics 188:151–164.
\bibAnnoteFile{wu2011joint}

\bibitem[{Yang et~al.(2023)Yang, Xu, Conant, Kishino, Thorne, and
  Ji}]{yang2023interlocus}
Yang, Y., T.~Xu, G.~Conant, H.~Kishino, J.~L. Thorne, and X.~Ji. 2023.
  Interlocus gene conversion, natural selection, and paralog homogenization.
  Molecular Biology and Evolution 40:msad198.
\bibAnnoteFile{yang2023interlocus}

\bibitem[{Yeang and Haussler(2007)}]{yeang2007detecting}
Yeang, C.-H. and D.~Haussler. 2007. Detecting coevolution in and among protein
  domains. PLoS computational biology 3:e211.
\bibAnnoteFile{yeang2007detecting}

\end{thebibliography}

\appendix
\setcounter{figure}{0}
\renewcommand{\thefigure}{S\arabic{figure}}

\section{Fixed-data benchmark figures}

The main results report time per MCMC iteration for benchmarks that vary the number of site patterns.
Here we show supporting fixed-data benchmarks that instead report total computation time for the same short MCMC analysis under each method and thread count.

\begin{figure}[H]
  \begin{centering}
  \includegraphics[width=\textwidth,height=0.5\textwidth,keepaspectratio]{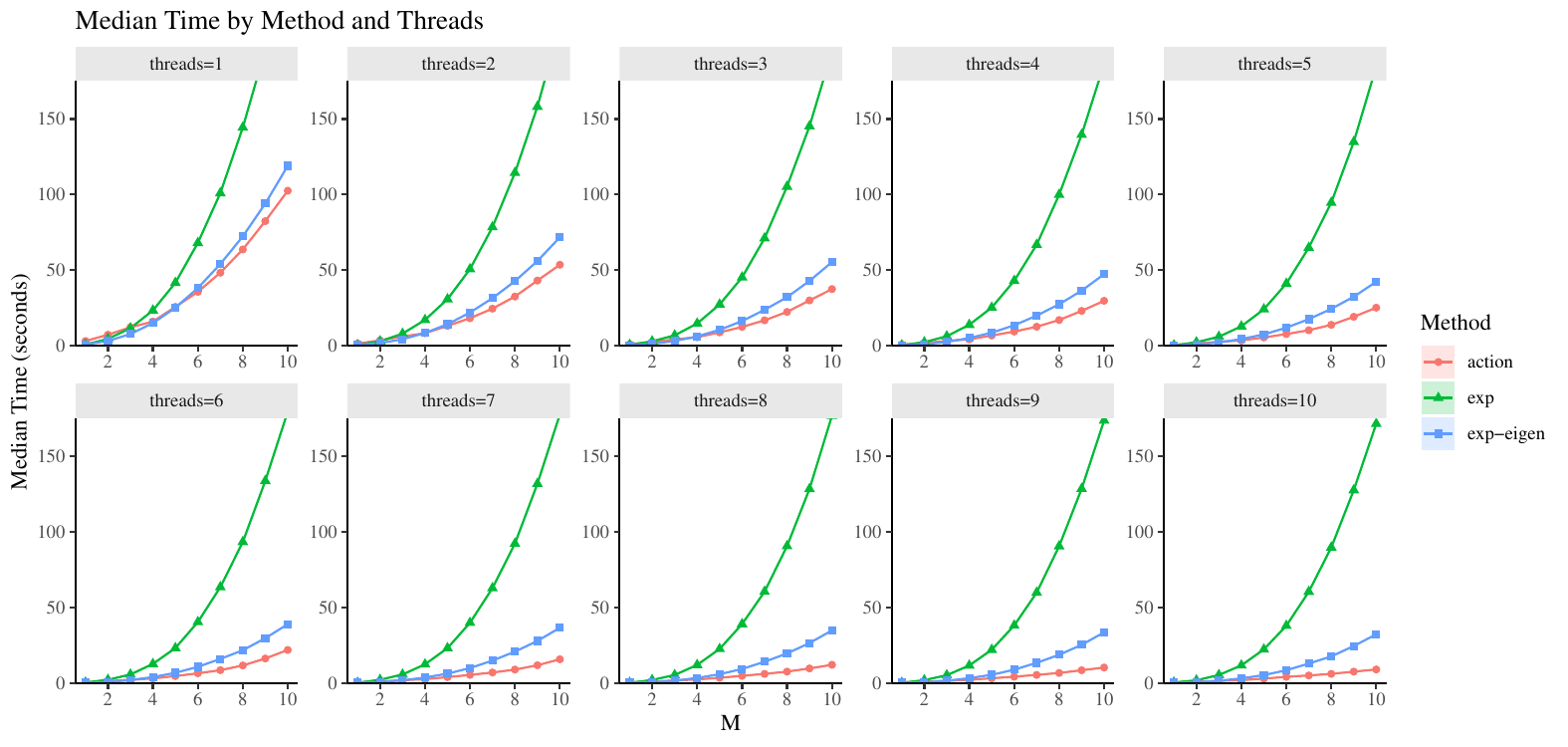}
  \par\end{centering}
\caption{\label{app:timeVsStates}Fixed-run Markov-modulated benchmark on the full 3,052-pattern data set, varying selective regimes and threads.
This reports total runtime for the same short MCMC analysis, not time per iteration.}
\end{figure}

\begin{figure}[H]
  \begin{centering}
  \includegraphics[width=\textwidth,height=0.6\textwidth,keepaspectratio]{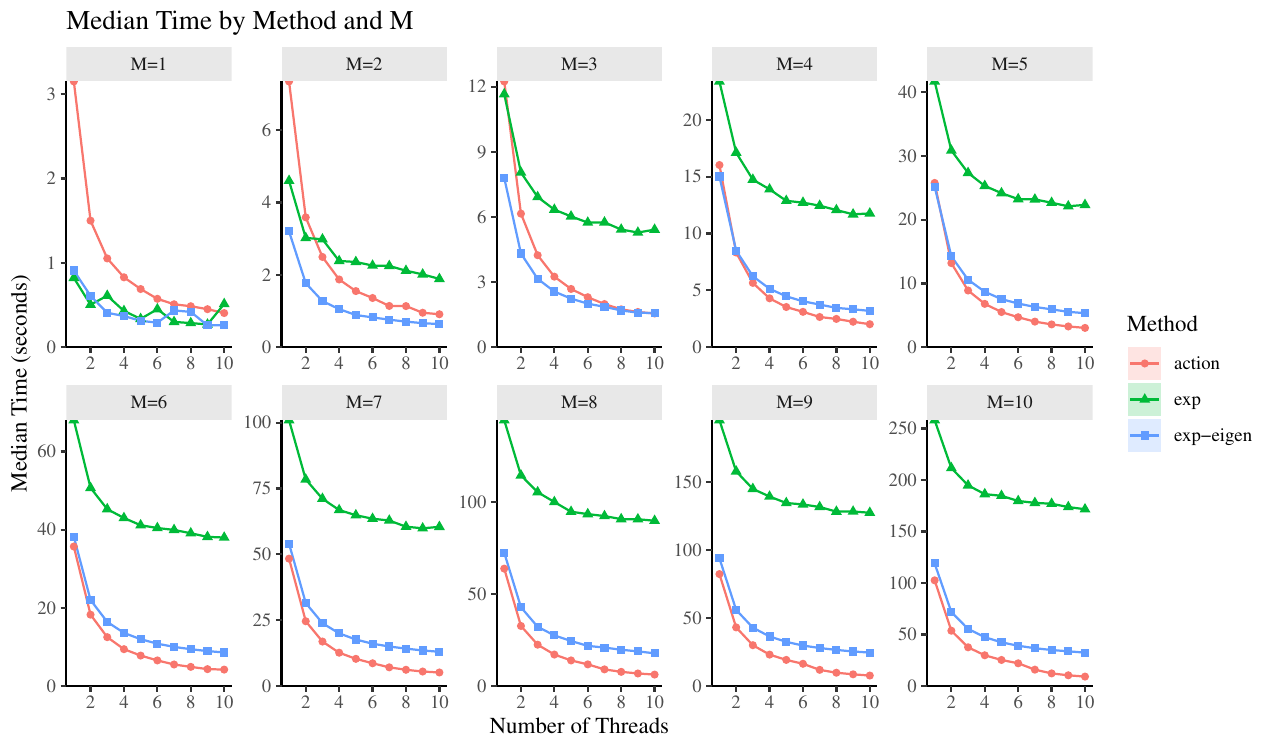}
  \par\end{centering}
\caption{\label{app:timeVsThreads}Fixed-run Markov-modulated threading benchmark on the full 3,052-pattern data set.
This figure reports total runtime for the same short MCMC analysis, not time per iteration.}
\end{figure}

\begin{figure}[H]
  \begin{centering}
  \includegraphics[height=0.45\textwidth]{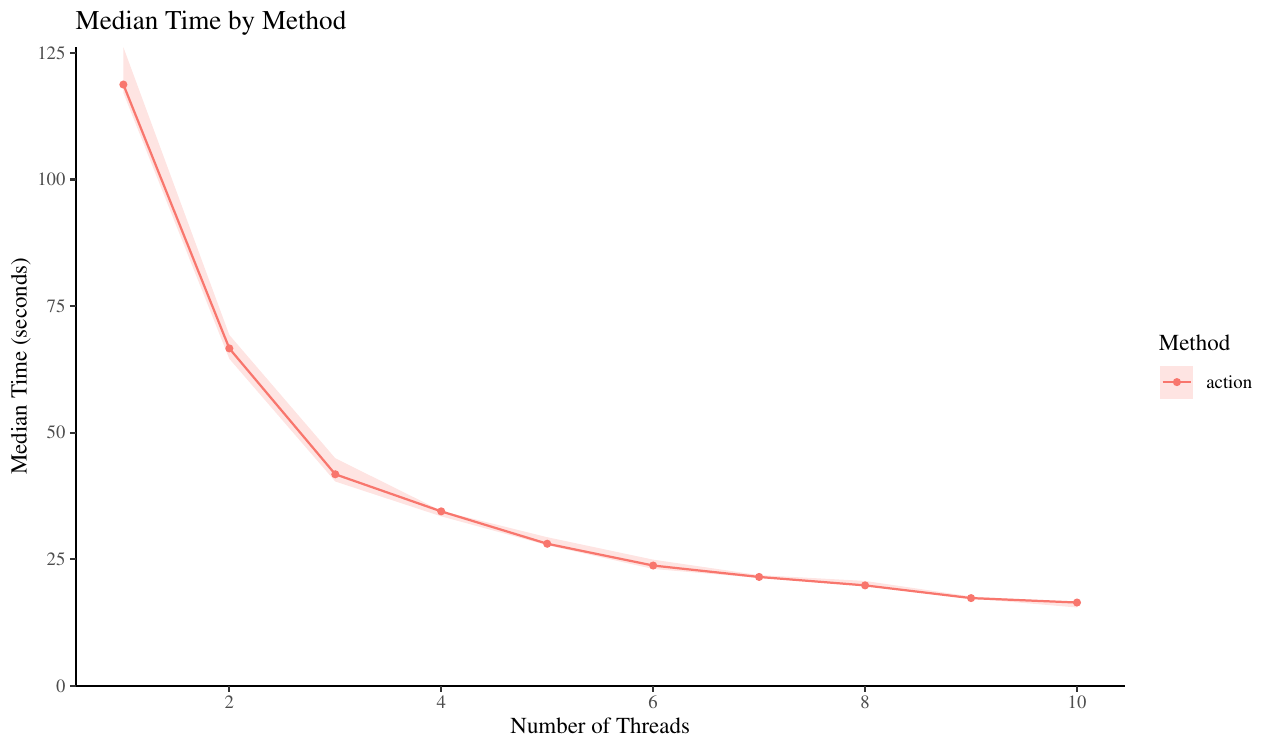}
  \par\end{centering}
\caption{\label{app:timeVsThreadsIGC}Fixed-run IGC threading benchmark on the original 90-pattern data set.
This figure reports total runtime for the same short MCMC analysis, not time per iteration.}
\end{figure}

\end{document}